\documentclass[twocolumn,showpacs]{revtex4}

\usepackage{amsmath,latexsym,amssymb}
\newcommand{\be}{\begin{equation}}
\newcommand{\ee}{\end{equation}}
\newcommand{\bea}{\begin{eqnarray}}
\newcommand{\ena}{\end{eqnarray}}

\renewcommand\l{\lambda}
\renewcommand\o{\omega}
\renewcommand\a{\alpha}

\renewcommand\l{\ensuremath{\lambda}}
\newcommand\s{\ensuremath{\sigma}}
\newcommand\x{\ensuremath{\times}}

\newcommand\m{\ensuremath{\mu}}
\renewcommand\k{\ensuremath{\kappa}}

\newcommand\n{\ensuremath{\nu}}
\newcommand\M{{\ensuremath{\cal M}}}

\newcommand\tr{\text{tr}}
\newcommand\diag{\text{diag}}
\newcommand\PM[1]{\begin{pmatrix}#1\end{pmatrix}}
\newcommand{\de}{\partial}

\newcommand{\ben}{\begin{displaymath}}
\newcommand{\een}{\end{displaymath}}
\newcommand{\ba}{\begin{eqnarray}}
\newcommand{\ea}{\end{eqnarray}}
\newcommand{\ban}{\begin{eqnarray*}}
\newcommand{\ean}{\end{eqnarray*}}
\newcommand{\bs}{\begin{split}}
\newcommand{\es}{\end{split}}

\begin{document}
\title{Spontaneous Lorentz Breaking and Massive Gravity}
\author{Z. Berezhiani$^a$, D. Comelli$^b$, F. Nesti$^a$ and   L. Pilo$^a$}
\affiliation{%
  $^a$Dipartimento di Fisica, Universit\`a di L'Aquila, I-67010 L'Aquila, and\\
  INFN, Laboratori Nazionali del Gran Sasso, I-67010 Assergi, Italy\\
  $^b$INFN - Sezione di Ferrara,  I-35131 Ferrara, Italy
}

\begin{abstract}
We study a theory where the presence of an extra spin-two field
coupled to gravity gives rise to a phase with spontaneously broken
Lorentz symmetry. In this phase gravity is massive, and the Weak
Equivalence Principle is respected. The newtonian potentials are in
general modified, but we identify an non-perturbative symmetry that
protects them.  The gravitational waves sector has a rich
phenomenology: sources emit a combination of massless and massive
gravitons that propagate with distinct velocities and also oscillate.
Since their velocities differ from the speed of light, the time of
flight difference between gravitons and photons from a common source
could be measured.
\end{abstract}


%\date{\today}% : \textbf{\jobname}}

\maketitle

%\section{Intro}
Since the early days of General Relativity (GR) there has been
attempts to modify gravity at long distances. The problem is not just
a theoretical challenge, but it could bring to important
phenomenological consequences. The only ghost free massive deformation
of linearized GR in the Minkowski background was given in~\cite{PF} by
Pauli and Fierz~(PF).  The peculiarity of gravity became apparent when
it was realized that in the zero mass limit the PF theory was
discontinuous~\cite{DIS} due the presence of an extra polarization
state.  The mystery of massive gravity deepened when it was realized
that the propagation of five degrees of freedom is spoiled by
interactions and a sixth ghost-like state turns on~\cite{BD}.  The
problem was reexamined in the framework of effective field theory
realizing that the reason behind the misbehavior of PF massive gravity
was strong coupling~\cite{NGS}.  It remains an open question whether a
well behaved non-linear extension of the PF theory exists (see for
instance \cite{DAM}) but Lorentz invariance clearly plays a crucial
restrictive role. If one gives up Lorentz invariance a trouble-free
massive deformation of gravity where the strong coupling and
discontinuity issues are disentangled can be found~\cite{RUB,DUB}.

In this letter we describe a mechanism of spontaneous lorentz breaking
that provides mass to the graviton in a consistent way.

Let us consider a theory with two dynamical metrics $g_1{}_{\m\n}$,
$g_2{}_{\m\n}$ \footnote{A spin two field of non geometrical origin
  may appear as fluctuations around a tensor condensate,
  Z. Berezhiani, O. Kancheli, in preparation.}  each interacting with
its own matter, a so called bigravity theory. The action contains two
Einstein-Hilbert (EH) terms, and a mixed term~\cite{Isham}:
\vspace*{-1ex}
\bea
S\!\!& =&\!\!
\int \!\!d^4 x\,  \Big[ \sqrt{g_1} \left(M_{1}^2 R_1+ {\cal L}_{1} \right) + \sqrt{g_2} \left( M_{2}^2 R_2+{\cal L}_{2}\right)  \nonumber\\[-.5ex]
&&{} \qquad- 4  (g_1 g_2)^{1/4}  V(g_1,g_2) \Big]\,,
\nonumber\\[-4ex]
\ena
where ${\cal L}_{1,2}$ are the corresponding matter lagrangians. 

The mixed term $V$ contains only non-derivative couplings between two
metrics, therefore it can only be function of the tensor $X^\mu_\nu =
{g_1}^{\mu \alpha} {g_2}_{\alpha \nu}$\cite{DAM1}.  The cosmological terms can be
included in $V$, e.g.\ $V_{\Lambda_1} = \Lambda_1 q^{-1/4}$, with $q =
\det X=(g_2/g_1)^{1/4}$.  The equations of motion (EoM) are
\vspace*{-2ex}
\be
\begin{split}
M_{1}^2 \, {E_1}_\mu^\nu +q^{1/4}
\left( V \delta_\mu^\nu -
 4 {V^\prime}_\mu^\alpha X_\alpha^\nu \right) =\frac{1}{2}{T_1}_\mu^\nu& 
\\
M_{2}^2 \, {E_2}_\mu^\nu + q^{-1/4}
\left( V \delta_\mu^\nu +
 4 {V^\prime}^\nu_\alpha X^\alpha_\mu \right) =\frac{1}{2}{T_2}_\mu^\nu &,
\\[-1ex]
\end{split}
\label{eq:eom}
\ee
where ${V^\prime}_\m^\n$ is the derivative of $V$ with respect to
$X^\m_\n$. 

\pagebreak[3]
\noindent  The indices of the two equations are
raised/lowered with the corresponding metrics.  

The action is invariant under a generic infinitesimal diffeomorphism
(diff) generated by $\xi^\mu$:
\be
\label{eq:diff}
\delta g_a{}_{\mu \nu} = \de_\mu \xi^\alpha g_a{}_{\alpha \nu} +
\de_\nu \xi^\alpha g_a{}_{\mu\alpha} + \xi^\alpha \de_\alpha
g_a{}_{\mu \nu}\,,
\ee
with $a=1,2$. Notice that in the absence of $V$ the system has a
larger gauge symmetry: one can use different $\xi^\mu_{1}$ and
$\xi^\mu_{2}$ for $g_1$ and $g_2$.  The ``diagonal'' diff invariance is
encoded in a set of generalized Bianchi identities.

The two metrics can be diagonalized simultaneously, but in general
their eigenvalues will not be proportional, and thus local Lorentz
invariance will be broken.  For vacuum solutions, we will assume
that rotational invariance is preserved and that the two metrics have
the same signature.

\medskip

\noindent {\bf\slshape The vacuum.}  
The EoM~(\ref{eq:eom}) always admit constant curvature solutions for
both $g_1$ and $g_2$, and in addition curvatures are
proportional~\cite{US}. For simplicity here we focus only on the flat
limit of those solutions, for which one fine tuning on $V$ is
necessary, as in standard GR for setting the cosmological term to
zero. In this biflat case, the EoM are simply
\begin{gather}
\label{bc}
\bar V=0\,,\qquad \bar V^\prime{}_\m^\n=0\,,
\end{gather}
where the bar stands for the background values.

For rotationally invariant backgrounds, these are three independent
equations.  One of the equations corresponds to the mentioned
fine-tuning for biflat backgrounds.  Then assuming that we live in
sector 1 we can set
\vspace*{-1ex}
\be
\begin{split}
& {\bar g}_{1\,\mu\nu} = \diag(-1,1,1,1) \\[.5ex]
& {\bar g}_{2\,\mu\nu} = \omega^2\,\diag(-c^2,1,1,1)\,.
%
%ds_i^2 = \omega^{2(i-1)}
%(- c^{2(i-1)}\, dt^2 + d\vec{x}^2) \, ,
%i=1,2\nonumber
\label{vac}
\end{split}
\ee
Thus, the two remaining equations determine the constants $c$ and
$\omega$ for any given $V$.  Physically $c$ is the speed of light in
sector 2, while $\omega$ parametrizes the relative conformal factor.
However, a solution with $c=1$ is always present, since in this case
two equations coincide (and determine $\o$).

Summarizing, we have two branches of solutions: Lorentz Invariant (LI)
for $c=1$, and Lorentz Breaking (LB) for $c\neq1$~\footnote{The
special case when $V=V(\det X)$ is exactly solvable: Bianchi
identities force $\det X$ to be constant, and there are additional
gauge symmetries that allow to set $c=1$. As a result both branches
are equivalent~\cite{US}.}.

The LB branch is of particular interest since it naturally allows for
consistent massive deformations of gravity.

\pagebreak[3]

\medskip

\noindent
{\bf\slshape Linearized analysis.}
Let us consider perturbations around the background (\ref{vac}),
defined as $g_{1\m\n} = {\bar g}_{1\m\n} + h_{1\m\n}$, and $g_{2\m\n}
= {\bar g}_{2\m\n} + \o^2 h_{2\m\n}$.  The
diffeomorphisms~(\ref{eq:diff}) act at lowest order as $\delta
h_{1\,\m\n}=2 {\bar g}_{1\a(\m}\de_{\n)} \xi^{\a}$ and $\delta
h_{2\,\m\n}=2\o^{-2} {\bar g}_{2\a(\m}\de_{\nu)} \xi^{\a}$.  Since
the background preserves rotations, we decompose both perturbations
according to their spin content:
\vspace*{-.5ex}
\be
\begin{split}
& {h_a}_{00} = \psi_a \,, \qquad {h_a}_{0 i} = {u_a}_i + \de_i v_a\,, \\[.5ex]
&  {h_a}_{ij} = {\chi_a}_{ij} + \de_i {S_a}_j + \de_j {S_a}_i + \de_i \de_j \sigma_a + 
\delta_{ij} \, \tau_a \,,\\[-.5ex]
\end{split}
\ee
with $\de_i {u_a}_i= \de_i {S_a}_i = \de_j {\chi_a}_{ij} =
\delta_{ij}{\chi_a}_{ij} = 0$.  We have thus 1+1 gauge invariant
tensors, 2+2 vectors, 4+4 scalars. Also the 4 diagonal diffs can be
split into one vector $\xi^i_T$ and 2 scalars $\xi^0$, $\de_i
\xi^i$. As a result, we expect that 2 tensors, 3 vectors and 6 scalars
are physical states determined by the EoM, while the remaining vector
and 2 scalars can be gauged away.

At quadratic level, in general the Lagrangian has the form $L =
L_{kin} + L_{mass}+ L_{source}$, where $L_{kin}$ contains derivative
terms (in space and time) emerging from the EH actions, $L_{mass}$
comes the the quadratic expansion of the mixed term, while
$L_{source}$ describes the gravitational coupling with matter.  In
field space, each field is a 2-component column vector, and it is
convenient to define the following 2\x2 matrices: $C = \diag (1, c) $,
$M^2 = M_1^2\diag (1,\k)$ where $\k= M_{2}^2/M_1^2\o^2c$. Then the
kinetic term has the compact form
\vspace*{-1ex}
\ba
\label{kin}
L_{kin}\!&=&\!
\frac{1}{4} \chi_{ij}^t M^2  \left( C^2 \Delta - \de_t^2 \right)\chi_{ij}
-\frac12 W_i^t M^2 \Delta  W_i^t+
\nonumber \\
\!&&\!+\frac12\Big[ 2 \Phi^t M^2    \Delta \,   \tau - \tau^t M^2 \left(C^2 \Delta -3 \de_t^2  \right) \tau\Big],  
\\[-3ex]
\nonumber
\ea 
where $W_i = u_i - \de_t S_i $, $\Phi = \psi - 2 \de_t v + \de_t^2
\sigma$.  Notice that $\chi_{ij}$, $W_i$, $\Phi$, $\tau$ are all gauge
invariant fields making the invariance of $L_{kin}$ manifest.  When
the matter energy-momentum tensors are conserved also the source
lagrangian is expressed in terms of gauge invariant fields:
\vspace*{-3ex}
\ba
\label{source}
L_{source} \!&=&\! 
\frac12\Big(\!-T^t_{ij} C \chi_{ij} + 2 T^t_{0i} \Omega C^{-1} W_i +\nonumber\\
\!\!&&\quad{}-  T^t_{ii} C \tau - T^t_{00} C^{-3} \Phi\Big)\,.
\ea

The crucial term, $L_{mass}$, in the flat limit stems from the
expansion of $V$. For the rotational invariant biflat background
({\ref{vac}), for which $\bar V=\bar V'=0$, it reads
\vspace*{-.5ex}
\ba
\label{mass}
 L_{mass}\!&=&\! 
-2  \left( {\bar g}_1 {\bar g}_2 \right)^{1/4} 
 \, \text{Tr} \left(X_{1} \, \bar { V}'' \, X_{1} \right) \nonumber\\
 \!&\equiv&\! \frac14\Big( \M_0^{ab} h_{00}^a  h_{00}^b
     +2 \M_1^{ab} h_{0i}^a h_{0i}^b  - \M_2^{ab} h_{ij}^a  h_{ij}^b\nonumber\\ 
  \!&&\!{}  + \M_3^{ab} h_{ii}^a  h_{ii}^b
     - 2  \M_4^{ab} h_{00}^a h_{ii}^b \Big)\,, \nonumber\\[-4ex]
\ea
where $X_{1} = \bar X {\bar g}_2^{-1} h_2 - {\bar g}_1^{-1} h_1 \,
\bar X$ is the fluctuation of $X$.  The second line defines the
Lorentz breaking masses $\M$~\cite{RUB} that can be computed
explicitly once a $V$ and a consistent background are given. Their
pattern is drastically different in the two branches.

Gauge invariance (\ref{eq:diff}) gives crucial constraints:
\vspace*{-1ex}\bea
\label{condM}
&\M_{0,1 }\begin{pmatrix}1 \\ c^2   \end{pmatrix}=0\,,
 \qquad\M_{1,2,3,4} \begin{pmatrix} 1\\ 1\end{pmatrix}=0\,,
\nonumber\\
& 
\begin{pmatrix} 1 & c^2 \end{pmatrix} \M_4 =0 \,.
\ena
In the LB phase, $c \neq 1$, this implies that $\M_1$=0 and
\ba
&\M_0 = \lambda_0  \begin{pmatrix} 1 & -\frac{1}{c^2} \\
  -\frac{1}{c^2} & \frac{1}{c^4}  \end{pmatrix} \qquad 
\M_4 = \lambda_4 \, \begin{pmatrix} 1 & -\frac{1}{c^2} \\
  -1 & \frac{1}{c^2}  \end{pmatrix},
\nonumber\\[1ex] 
\label{eq:masses}
&\M_{2,3}=\lambda_{2,3} \, \begin{pmatrix} \,1 & -1 \\ -1\, & 1  \end{pmatrix}.
\ea
In the LI phase instead $c=1$ and two conditions in~(\ref{condM})
coincide, thus allowing a non-vanishing $\M_1$. Also, all the $\M$'s
are proportional to the same projector (\ref{eq:masses}).  Moreover,
in this branch the mass term reduces to the generalized PF form:
$L_{mass}= \eta^{\mu \alpha} \eta^{\nu \beta} h^t_{\mu \nu} A
h_{\alpha \beta} + h^t B h$ ($A$ and $B$ being 2$\times$2
matrices). This is equivalent to $\M_0=A+B$, $\M_{1,2}= -A$,
$\M_{3,4}=B$. We remark that the limit $c \to 1$ is discontinuous.

Let us analyze in detail the linearized theory, separately in the two
branches.

\begin{table*}
\def\arraystretch{1.2}
\newcommand\MC[2]{\multicolumn{#1}{l|}{#2}}
\newcommand\MCc[2]{\multicolumn{#1}{c|}{#2}}
\newcommand\MCcc[2]{\multicolumn{#1}{c}{#2}}
\newcommand\B[1]{{\fboxsep=1.7pt\fbox{$#1$}}}

\begin{tabular}{l|l|c|c|c|c|c|c|c|l|c}
&\MCc{1}{Phase}        &   $\l_0$ & $\l_1$ & $\l_2$ & $\l_3$ & $\l_4$ & Extra Gauge  & \MCc{1}{States} & \MCc{1}{Static Potentials}  & $V(X^\m_\n)$ \\[.1em]
\hline                
\hline             
&PFGeneric             & $a\!+\!b$ & $-a$ & $-a$ & $b$ & $b$ & - & ghost  & & Generic \\
\MC{1}{$c=1$}
&PF                    &   $0$ &  $b$ &  $b$ & $b$ & $b$ & - & \B{2_0}+\B{2_m}  & GR$_+$ + GRdisc$_-$& Fine tuned\\
&PF0                   &   $b$ &  $0$ &  $0$ & $b$ & $b$ & 3 transverse diffs & 2\x\B{2_0}  & 2\x GR & $V(\det X)$\\
\hline	              
%$c\not=1$             &       &      &      &     &     &   &     &      \\
\hline             
&Generic                 &   *   &   0  &  *   & *   &  *  & - & \B{2_0}+\B{2_m} & GR$_+$ + GR$\m_-$& Generic \\
&$\l_0=\l_4=0$            &   0   &   0  &  *   & *   &  0    & $\psi_-$ & \B{2_0}+\B{2_m} & GR$_+$ + GR$\m_-$  \hfill($^1$)\\
&$\l_\mu=0$               &   *   &   0  &  *   & *   &  *    &   -     & \B{2_0}+\B{2_m} & 2\x GR \hfill ($^2$)\\
\MC{1}{$c\not=1$}&
$\l_\m=0$, $\l_0=-3\l_4$ & $-3b$ &   0  &  $a\!+\!b$   & $a$   &  $b$  &  - &  \B{2_0}+\B{2_m}  & 2\x GR \hfill\ \  (Weyl$_-$) ($^2$) & Homogeneous\\
&$\l_\eta=0$              &   *   &   0  &  *   & *   &  *  & $\psi_-\!+\!\s_-\!+\!\Phi_u$ &  \B{2_0}+\B{2_m} & GR$_d$\hfill ($^3$) \\
&$\l_\m,\l_\eta,\l_2=0$   &   *   &   0  &  0   & *   &  *  &  3 transverse diffs   &  2\x\B{2_0}  & 2\x GR  & $V(\det X)$\\
&$\l_0\not=0$, $\l_i=0$  &   *   &   0  &  0   & 0   &  0  &  3 spatial diffs  &  2\x\B{2_0}  & 2\x GR  & \\
\hline
\end{tabular}
\caption{Summary of phases in the biflat limit. Propagating states are labeled as \B{2_0}, \B{2_m} for spin 2 massless, massive. 
  GR$_{\pm}$ refers to standard GR potentials for the $\pm$ combinations of $\Phi$, $\tau$.  GRdisc refers to GR potentials plus discontinuity,   
  GR$\m$ refers to  GR potentials plus linear term (\ref{eq:potlin}).
  ($^1$)~Here $\m^2=\l_2(3\l_3-\l_2)/(\l_3-\l_2)M_1^2$.
  ($^2$)~A flat direction of the $V$ potential, 
  $\delta(\psi_-,\tau_-)\propto(3\l_4,\l_0)$, or Weyl$_-$ for $3\l_4=-\l_0$: 
  these guarantee the vanishing of the extra linear potential.
  ($^3$)~The source $T_{00}^-$ of the undetermined field $\Phi_u$ should vanish;
  the orthogonal combination $\Phi_d$ has Newtonian potentials.
}
\label{tab:phases}
\end{table*}

\medskip
\noindent \textbf{Lorentz invariant phase.} 
Since here all the mass matrices are proportional to the same
projector with a null eigenvalue, the linearized theory is
diagonalized by the mass eigenstates
\be
\begin{split}
&h^{c+}_{\m\n}= \cos\theta \,h^{c1}_{\m\n} + \sin\theta \,h^{c2}_{\m\n}\\
&h^{c-}_{\m\n}= \sin\theta \,h^{c1}_{\m\n} - \cos\theta \,h^{c2}_{\m\n}\,,
\end{split}
\ee where $h^{c1,2}$ are the canonically normalized fields and the
mixing angle is $\sin\theta=(1+\k)^{-1/2}$. The state $h^{c+}$ is
massless, while $h^{c-}$ has a generalized PF mass term.

The massive sector $h^{c-}$ will be plagued by the known problems
related to the Pauli-Fierz case.  In the general case, with $A\neq-B$,
it contains ghosts~\footnote{Ghosts are absent in the LI phase also
  with $A=0$, but the resulting theory is rather boring:
  there are additional gauge symmetries and the spin 2 states stay
  massless and decoupled.}.
If we remove the ghosts setting $A=-B$, we face the vDVZ discontinuity
problem~\cite{DIS}.  In this theory however, since the original
gravitons have different Planck scales, the discontinuity in our
sector can be controlled by the mixing angle.  The present
bounds~\cite{Will} would require $M_{2}\sim 30 M_{1}$.  In addition
the Newton constant becomes distance dependent: it changes by a factor
of $1+4\tan^2\theta/3$ from short to large distance, the critical
distance being the inverse graviton mass.  The matter of type 2, if it
exists, will interact with ours in distance dependent
way~\cite{ROSSI}.

Finally, the sixth mode \cite{BD} of $h^{c-}$ will manifest once 
interactions are considered, and in general will propagate in both
gravities, leading to strong coupling~\cite{NGS}.

\medskip
\noindent \textbf{Lorentz breaking phase.} 
We give an overall analysis of the static and propagating modes in the
LB branch, and refer to table~\ref{tab:phases} for the full summary of
all phases; details will be given in~\cite{US}.

\paragraph{Tensors} 
Each tensor has two independent components.  They propagate according
to the EoM:
\be \left[ \PM{(k_0^2-\vec k^2)& 0\\ 0& \k(k_0^2-c^2\,\vec k^2) } -
  \frac{\M_2}{M_1^2} \right]\chi_{ij}=0\,,  \ee
describing two gravitons with different ``speeds of light''.  The
gravitons are mixed by the non-diagonal mass matrix $\M_2$, and thus
oscillate and have a nonlinear dispersion relation.  Since $\M_2$ has
a zero eigenvalue, expanding the dispersion relation in powers of
$k^2$ one finds at low energy a massless graviton that travels with
speed $v^2=(1 + c^2 \k) / (1+\k)$, and a massive one, of mass
$m_g^2=(1 + \k^{-1})\l_2/M_1^2$. In the high energy limit two states
propagate with different speeds and also oscillate.

Interestingly, when $c>1$, the second graviton propagates faster than
``our'' light, though this will not lead to causality violations.
Indeed, in the coordinates where the metrics are~(\ref{vac}), the
Cauchy problem is globally well posed in terms of the preferred time
$t$.  In principle this scenario could be tested by observing the time
of flight difference between gravitational waves and optical signals,
or frame dependence of the gravitational waves propagation that would
provide the evidence for a preferred frame.

\paragraph{Vectors}

Thanks to $\M_1=0$, the vector states do not propagate.  The EoM
however determine three static potentials: the two $W_{1i}$, $W_{2i}$
that couple to the respective $T_{0i}$'s as in standard GR, without
mixing, and the combination $S^-_i=(S_{1i}-S_{2i})$, that is
zero.

\paragraph{Scalars} Also scalar modes do not
propagate, again because of $\M_1=0$; they however mediate the static
potentials.  By defining $\l_\eta^2 \equiv \l_4^2+ \l_0 (\l_2-\l_3)$,
$\l_\mu^2 \equiv 3 \l_4^2+ \l_0 (\l_2 - 3 \l_3)$ and
$T_{00}^-\equiv({T_1}_{00} - {T_2}_{00}/c^4\o^2\k)$, the solutions of
the EoM are
\ba
\label{eq:potlin}
\Phi_1\!\!&=&\!\!\frac{1}{2 M_{1}^2\Delta }
              ({T_1}_{00}+{T_1}_{ii}-3\frac{{\ddot{T}{}_1}_{00}}{\Delta} )
+\frac1{M_1^2}\frac{\m^2}{\Delta^2}T_{00}^-
\nonumber\\
\Phi_2\!\!&=&\!\!\frac{c^{-1}}{2M_{2}^2\Delta}
              ({T_2}_{00}\!+\!c^2{T_2}_{ii}-\!3\frac{{\ddot{T}{}_2}_{00}}{c^2\Delta} )
-\frac{1}{M_2^2}\frac{\m^2}{\Delta^2}T_{00}^-
\nonumber\\
\tau_1\!\!&=&\!\!\frac{1}{2 \Delta}\frac{{T_1}_{00}}{M_{1}^2}\,,\qquad
\tau_2=\frac{c^{-3}}{2 \Delta}\frac{{T_2}_{00}}{M_{2}^2}\,.
\ea
where 
\be
\label{eq:mu}
\m^2=\frac{\l_2}{2M_1^2}\frac{\l_\m^2}{\l_\eta^2}\equiv\frac{\l_2}{2M_1^2}\frac{3\l_4^2-\l_0(3\l_3-\l_2)}{\l_4^2-\l_0(\l_3-\l_2)}\,,
\ee
The two remaining gauge invariant fields, $\psi_-\equiv
(c^2\psi_1-\psi_2)$ and $\sigma_-\equiv (\sigma_1-\sigma_2)$ are
determined in terms of $\tau_-\equiv (\tau_1-\tau_2)$. Since they have
no source, the explicit expression is omitted.  From $\Phi_1$
in~(\ref{eq:potlin}), $M_1$ is identified as the Planck mass,
$M_1^2=M_P^2=(8\pi G)^{-1}$.

Since in this phase only gravitons propagate, strong coupling
problems~\cite{NGS} are absent.

\medskip

\noindent
{\bf\slshape Linear term.}
In (\ref{eq:potlin}) we recognize the standard GR potentials plus, in
$\Phi_{1,2}$, a term proportional to $1/\Delta^2$, that represents a
linearly growing potential at large distances~\cite{DUBA}. For generic
$\l_{0,2,3,4}$, the scale at which this force sets out is proportional
to $\m$, (\ref{eq:mu}). Since this linear growth signals the breakdown
of perturbation theory at large distances from sources, it is
interesting to study the conditions under which $\m$ vanishes.  One,
cheap, possibility is to have massless gravitons, $\l_2=0$; the other
possibility is that the numerator vanishes,
$\l_\m^2=3\l_4^2-\l_0(3\l_3-\l_2)=0$.  The latter condition can be
understood as a symmetry requirement on the potential $V$ by
expressing it as a function of the fluctuation $X_{1}$: the most
general potential preserving rotations, with vanishing $\bar V$ and
$\bar V'$, can be written directly in terms of the lorentz breaking
masses~$\l_i$:
\bea
\nonumber\label{potV}
&&\!\!\!\!\!\!V=\l_0 \tr[X_{1}P_t]^2 - \l_2 \tr[X_{1}P_sX_{1}P_s]+ \l_3 \tr[X_{1}P_s]^2 +\\[.7ex]
&&\!\!\!\!+ 2 \l_4 \tr[X_{1}P_t] \tr[X_{1}P_s] + 2\l_1 \tr[X_{1}P_sX_{1}P_t]+\cdots
\ena 
with $P_t\!=\!\diag(1/c^2,0,0,0)$, $P_s\!=\!\diag(0,1,1,1)$.  In the
LB phase the last term is absent.  Requiring $\l_\mu=0$, we find that
the potential is invariant under
\be
\label{eq:mu0inv}
\delta X_{1}=\epsilon\,\text{diag}(3\l_4c^2,-\l_0,-\l_0,-\l_0)\,,
\ee
that shifts only the scalar fields $\psi_-$ and $\sigma_-$. This
transformation matches the one encountered for goldstone fields in the
effective description of~\cite{DUBA}. 

However, if we further restrict $3\l_4=-\l_0$ ($\gamma=1$
in~\cite{DUBA}) we find the ``anti-diagonal'' Weyl transformation
\be
\text{Weyl$_-$:}\qquad \delta X_{1}= \epsilon \bar X\quad \Leftrightarrow\quad   X\to (1+\epsilon)X\,,
\ee
generated by two opposite rescalings of $g_1$ and $g_2$.  We have thus
an explicit non-perturbative symmetry protecting $\m=0$, that gives
in addition $\l_0=-3\l_4$. Still the graviton mass $\l_2$ can be
nonzero and arbitrary. One example of a full potential invariant under
Weyl$_-$ is
$V=a_0+a_1\tr[X]\tr[X^{-1}]+a_2\tr[X^2]\tr[X^{-2}]$.

\pagebreak[3]

In the LI phase also there are interesting consequences: Weyl$_-$
makes all masses proportional, so that one cannot get rid of ghosts
($A+B=0$) without setting all masses to zero. This suggests that the
LB phase is the only physical one with massive gravitons.

It is interesting to note that $\m=0$ even when the potential is
homogeneous of a generic degree $\alpha\neq0$ under Weyl$_-$: $V(\l
X)=\l^\alpha V(X)$.  Interestingly enough, for ans\"atze of constant
curvatures this symmetry leads to a constraint on the curvatures as a
function of $\alpha$, with possible phenomenological consequences.
For instance, potentials of degree $\pm1$ lead to vanishing curvature
for $g_{1,2}$.  This opens up even the possibility to investigate the
stability of the LB phase under quantum corrections effects from
matter~\cite{US}.

\medskip

\noindent{\bf\slshape Phenomenology.}
First of all it is clear that since the lagrangian of our matter
${\cal L}_1$ is defined only with metric $g_1$, no lorentz breaking
effects can be observed in its propagation and interactions, and the
Weak Equivalence Principle is satisfied~\footnote{This would not be
  the case if one admits terms that couple our matter directly to
  $g_2$.}.  The same happens also for matter 2, though the speeds of
light in two sectors are different. Due to the interaction between
metrics, the Strong Equivalence Principle is violated in both sectors.

In the weak-field regime, besides the newtonian interaction $\propto
1/r$ we have an additional term that grows linearly with $r$.
Moreover, through this term the two kinds of matter see each other.
For instance, an ordinary source $M$ will produce potentials in both
sectors, $\Phi_1= - G M ( 1/ r + \m^2r)$ and $\Phi_2= + \k c\o^2 G M
\m^2 r $. Notice that the effect of the linear term is opposite in the
two sectors.

Of course if $\m=0$ as discussed above this effect is absent: both
potentials are newtonian, without mixing between the two sectors.  On
the other hand if $\m\neq 0$ the linear term generates a constant
acceleration in direction of the source, attractive or repulsive
depending on the sign of $\m^2$~\cite{DUBA}.  (Amusingly, this constant
acceleration can explain the Pioneer anomaly, $a\simeq10^{-9}\rm
m/s^2$, for $\m\sim 10^{-21}$eV).

\pagebreak[3]

The linearized solution can be trusted only from the Schwartschild
radius $r_{UV}=G M$ up to distances where the linear term drives
$\Phi$ to be $\sim1$; for instance, with matter of type 1, $r_{IR}= (G
M\m^2 )^{-1}$. In particular for the sun $r_{IR}\sim10^{12}{\rm
  pc}\times (\m/10^{-21}{\rm eV})^{-2}$, and clearly the linearized
approximation works fine for $\mu\sim 10^{-21}\rm eV$; on the other
hand, taking the galaxy as the source, $r_{IR}$ drops well below the
galactic radius. In such a situation a non-perturbative solution is
needed~\cite{US}.  We just observe here that since the curvature of
the linear term vanishes as $1/r$, one can hope to match with a
well-behaved solution at infinity.

Pulsar binary systems (BPS) set stringent limits on the rate of
gravitational waves (GWs) emission~\cite{Will}.  In our theory the
main effect is due to the fact that matter in sector 1 emits a
combination of graviton mass eigenstates, and the massive one is
forbidden if the energy is too low.  The emission rate will thus be
modified only if $m_g$ is higher than the GW energy \footnote{$m_g$ is
  linked to $\m$ by order one factors, $m_g\sim \m$, when the $\l$'s
  are all of the same order.}.  For BPS a rough limit is
$m_g<10^{-20}$eV.  However the BPS analysis involves the periastron
advance rate~\cite{Will} that is expected to be modified as well in
our case, therefore a complete study is needed.  Even if $m_g$ is
below such limit, there could be observable effects in GW detection:
as mentioned, the two graviton states have different speeds, therefore
it would be possible to observe a delay or anticipation of the GWs
with respect to optical signals coming from the same source.  Finally,
supposing that also type-2 matter can be a source of GWs, due to
oscillations one could see GW signals in our detectors not associated
with any otherwise visible sources like neutron stars etc.

We conclude that in this theory the spontaneous breaking of lorentz,
required to obtain an healthy massive gravity theory, turns into an
rich phenomenology, namely massive, oscillating and lorentz
breaking gravitational waves.

\medskip\noindent\textbf{\slshape Acknowledgments.}  Work partially
supported by the MIUR grant under the Projects of National Interest
PRIN 2004 ``Astroparticle Physics''.


\begin{thebibliography}{99}

\bibitem{PF} M.~Fierz, W.~Pauli,
%``On relativistic wave equations for particles of arbitrary spin in an
%electromagnetic field,''
\emph{Proc.\ Roy.\ Soc.\ Lond.\ } A {\bf 173}, 211 (1939).
%%CITATION = PRSLA,A173,211;%%

\bibitem{DIS}
H.~van Dam, M.J.G.~Veltman,
%``Massive And Massless Yang-Mills And Gravitational Fields,''
%%CITATION = NUPHA,B22,397;%%
\emph{Nucl.\ Phys.\ }  B {\bf 22} (1970) 397;
Y.~Iwasaki,
%``Consistency condition for propagators,''
\emph{Phys.\ Rev.\ }  D {\bf 2} (1970) 2255;
%%CITATION = PHRVA,D2,2255;%%
V.I.Zakharov, \emph{JETP Lett.} {\bf 12} (1971) 198.

\bibitem{BD}
D.G.~Boulware, S.~Deser,
%``Can gravitation have a finite range?,''
\emph{Phys.\ Rev.\ } D {\bf 6} (1972) 3368.
%%CITATION = PHRVA,D6,3368;%%

\bibitem{NGS}
N.~Arkani-Hamed, H.~Georgi, M.D.~Schwartz,
%``Effective field theory for massive gravitons and gravity in theory space,''
\emph{Annals Phys.\ } {\bf 305} (2003) 96.
%[arXiv:hep-th/0210184].
%%CITATION = APNYA,305,96;%%
\bibitem{Isham}
  C.~J.~Isham, A.~Salam and J.~A.~Strathdee,
  %``F-dominance of gravity,''
  Phys.\ Rev.\  D {\bf 3}, 867 (1971).
  %%CITATION = PHRVA,D3,867;%%
\bibitem{DAM}
T.~Damour, I.I.~Kogan, A.~Papazoglou,
%``Spherically symmetric spacetimes in massive gravity,''
\emph{Phys.\ Rev.\ } D {\bf 67} (2003) 064009; 
%[arXiv:hep-th/0212155].
%%CITATION = PHRVA,D67,064009;%%
%\bibitem{Dvali:2006su}
  G.~Dvali,
  %``Predictive power of strong coupling in theories with large distance
  %modified gravity,''
  New J.\ Phys.\  {\bf 8} (2006) 326.
%  [arXiv:hep-th/0610013].
  %%CITATION = NJOPF,8,326;%%

\bibitem{RUB}
  V.A.~Rubakov,
  %``Lorentz-violating graviton masses: Getting around ghosts, low strong
  %coupling scale and VDVZ discontinuity,''
  arXiv:hep-th/0407104.
  %%CITATION = HEP-TH/0407104;%%

\bibitem{DUB}
S.L.~Dubovsky,
%``Phases of massive gravity,''
\emph{JHEP} {\bf 0410}, 076 (2004).
%[arXiv:hep-th/0409124].
%%CITATION = JHEPA,0410,076;%%

\bibitem{DAM1}
  T.~Damour, I.I.~Kogan,
  %``Effective Lagrangians and universality classes of nonlinear bigravity,''
  Phys.\ Rev.\  D {\bf 66} (2002) 104024.
%  [arXiv:hep-th/0206042].
  %%CITATION = PHRVA,D66,104024;%%

\bibitem{US} Z. Berezhiani, D. Comelli, F. Nesti,
L. Pilo, in preparation.

\bibitem{DUBA} 
S.L.~Dubovsky, P.G.~Tinyakov, I.I.~Tkachev,
%``Cosmological attractors in massive gravity,''
\emph{Phys.\ Rev.\ } D {\bf 72}, 084011 (2005).
%[arXiv:hep-th/0504067].
%%CITATION = PHRVA,D72,084011;%%

%\cite{massivegrav}
\bibitem{massivegr}
  S.L.~Dubovsky, P.G.~Tinyakov, I.I.~Tkachev,
  %``Massive graviton as a testable cold dark matter candidate,''
  Phys.\ Rev.\ Lett.\  {\bf 94} (2005) 181102.
%  [arXiv:hep-th/0411158].
  %%CITATION = PRLTA,94,181102;%%

%\cite{Will:2005va}
\bibitem{Will}
  C.M.~Will,
  %``The confrontation between general relativity and experiment,''
  arXiv:gr-qc/0510072.
  %%CITATION = GR-QC/0510072;%%

%\bibitem{deffayet}
%  D.~Blas, C.~Deffayet and J.~Garriga,
%  %``Causal structure of bigravity solutions,''
%  Class.\ Quant.\ Grav.\  {\bf 23} (2006) 1697
%%  [arXiv:hep-th/0508163].
%  %%CITATION = CQGRD,23,1697;%%

\bibitem{ROSSI}
N. Rossi, in preparation.


\end{thebibliography}
\end{document}